\documentstyle[12pt,epsf,aaspp4]{article}
\font\smcap=cmcsc10
\def\farcs{\hbox{$\> .\!\!^{\prime\prime}$}}
\def\fdeg{\hbox{$\> .\!\!^{\circ}$}}
\def\fss{\hbox{$\> .\!\!^{\rm s}$}}
\def\refpar{\par\hangindent=3em\hangafter=1}
\def\reference{\relax\refpar}

\begin{document}

\singlespace

\title{Globular Cluster Photometry with the Hubble Space Telescope. \\
       VI.~WF/PC-I Observations of the Stellar Populations \\
       in the Core of M13 (NGC~6205)}

\author{Randi~L.~Cohen and Puragra Guhathakurta}
\affil{UCO/Lick Observatory, University of California, Santa Cruz, CA 95064,
USA}
\affil{Email: {\tt randi@ucolick.org}, \tt raja@ucolick.org}

\author{Brian Yanny}
\affil{Fermi National Accelerator Laboratory, Batavia, IL 60510, USA}
\affil{Email: \tt yanny@sdss.fnal.gov}

\author{Donald P.\ Schneider}
\affil{Department of Astronomy \& Astrophysics, The Pennsylvania State
	   University, \\
           University Park, PA 16802, USA}
\affil{Email: \tt dps@astro.psu.edu}

\author{John N.\ Bahcall}
\affil{Institute for Advanced Study, Princeton, NJ 08540, USA}
\affil{Email: \tt jnb@sns.ias.edu}

\def\reference{\relax\refpar}
\pagestyle{plain}

\newcommand{\E }{\eqno }
\newcommand{\I}{\indent}
\newcommand{\N}{\noindent}

\begin{abstract}

We study the dense core of the globular cluster Messier~13 (NGC~6205) using
pre-refurbishment Planetary Camera-I images obtained with the {\it Hubble
Space Telescope}.  Short exposures (60~s) through the F555W and F785LP
filters (similar to Johnson $V$ and $I$, respectively) have been used to
obtain $V$ and $I$ photometry of 2877~stars brighter than $V\sim20$ in a
1.25~arcmin$^2$ region of the cluster including its core and extending out to
$r\sim66^{\prime\prime}$ (2.3~pc) from its center.  The sample is complete
to $V\simeq18.3$ (the main sequence turnoff) and the $1\sigma$ photometric
error is about 0.1~mag.  We find 15~blue straggler star
candidates and 10~other possible blue stragglers in this region of M13.
Their specific frequency is in the range $F_{\rm BSS}=0.04$--0.07, comparable
to what is observed near the centers of other dense clusters.  A comparison
between M13's observed $V$ band stellar luminosity function and a theoretical
model (Bergbusch \& Vandenberg 1992) for the luminosity function of an old,
metal-poor cluster shows that the model predicts too few of the brightest red
giants ($V\sim12.5$--15) by a factor of two relative to subgiants/turnoff
stars ($>6\sigma$ effect).  The radial distributions of the red giants, blue
stragglers, and subgiants are consistent with one another, and are well fit
by a King profile of core radius $r_{\rm
core}=38^{\prime\prime}\pm6^{\prime\prime}$ (90\% confidence limits) or
1.3~pc.  Stars in the blue horizontal branch of M13, however, appear to be
centrally depleted relative to other stellar types.

We combine data from three dense `King model clusters', M13, M3, and 47~Tuc,
and two post core collapse clusters, M30 and M15, and compare the
distributions of various stellar types as a function of
$(r/r_{\rm half~light})$ and $(r/r_{\rm core})$.  The horizontal branch stars
in the combined sample appear to be centrally depleted relative to the giants
(97\% significance)---this depletion is only a $1\sigma$--$2\sigma$ effect in
each of the clusters taken individually.  The blue stragglers in the
combined sample are centrally concentrated relative to the giants.

\end{abstract}


\section{INTRODUCTION}

This is the sixth in a series of papers describing {\it Hubble Space
Telescope\/} ({\it HST\/}) observations of the centers of the nearest
Galactic globular clusters with $|b|>15^\circ$.  The main scientific goals of
this program are to measure the shape of stellar density profile in clusters
and to understand the nature of evolved stellar populations in very dense
regions by probing the variation in the mix of stellar types as a function of
radius (and hence stellar density).  In this paper, we apply the techniques
developed in our earlier Planetary Camera-I (PC) studies of 47~Tuc
(Guhathakurta et~al.\ 1992, hereafter referred to as Paper~I) and M15 (Yanny
et~al.\ 1994a, hereafter referred to as Paper~II) to analyze data on M13
(NGC~6205).

The globular cluster M13 is the least dense of the clusters analyzed thus
far in this series of papers: M3, 47~Tuc, M30, and M15.  The core of M13 has
a projected stellar density of $\sim0.3$~arcsec$^{-2}$ (300~pc$^{-2}$) for
stars with $V<V_{\rm HB}+2$.  [The easy detectability of this sample of
bright stars (upto 2~mag fainter than the horizontal branch) makes it a
convenient tracer of the relative surface density of clusters (Bolte et~al.\
1993).]~~ By contrast, the density of M3's core is $\sim1.1$~arcsec$^{-2}$
(460~pc$^{-2}$), the density of 47~Tuc's core is $\sim3$~arcsec$^{-2}$
(6000~pc$^{-2}$), and the density in the central few arcseconds of M15 and
M30 exceeds 30~arcsec$^{-2}$ ($>10^4$~pc$^{-2}$).  Thus, the effect of
crowding on the completeness and photometric accuracy of our M13 sample is
less severe than in our earlier PC studies.

The total absolute visual magnitude of M13 is $M_V=-8.51$, based on an
estimated distance of 7.2~kpc (Djorgovski 1993).  The cluster has a
metallicity of $\rm [Fe/H]=-1.65$ (Djorgovski 1993) and has an extended blue
horizontal branch.  In fact, M13's horizontal branch is entirely on the blue
side of the RR~Lyrae instability strip and is significantly bluer than the
horizontal branches of other clusters of comparable metallicity (M3 and
NGC~7006), making it important in the search for the (unknown) second
parameter that is thought to control horizontal branch morphology in globular
clusters (Zinn 1986).  The surface brightness profile of M13 is well fit by a
King profile with a core of radius $r_{\rm
core}=53^{\prime\prime}\pm7^{\prime\prime}$ or 1.8~pc (Trager et~al.\ 1993).
The cluster displays a significant amount of rotation with a maximum rotation
speed of about 5$\,$km~s$^{-1}$ (Lupton et~al.\ 1987).

Its relative proximity to the Sun and large size have made M13 the subject of
several studies in the last four decades (Savedoff 1956; Baum et~al.\ 1959;
Sandage 1970; Cudworth \& Monet 1979; Lupton \& Gunn 1986; Laget et~al.\
1992; Guarnieri et~al.\ 1993).  Recently, Stetson (1996) derived an accurate
color--magnitude diagram of stars located just outside the cluster core
using images obtained with the Canada--France--Hawaii telescope.  Our {\it
HST\/} investigation of M13's core is complementary to Stetson's study.
Despite the broad and complicated wings of the pre-refurbishment {\it
HST\/}'s aberrated point spread function, its $0\farcs1$ (FWHM) core makes it
possible to detect faint stars in crowded fields.  Pre-refurbishment {\it
HST\/} data yield cleaner and more complete stellar samples in dense globular
cluster cores than even $0\farcs5$-seeing ground-based images (cf.~Bolte
et~al.\ 1993; Guhathakurta et~al.\ 1994, hereafter referred to as Paper~III).

In Sec.~2, we describe the acquisition and processing of the {\it HST\/} PC
data used in this study.  In Sec.~3, we present a color--magnitude diagram
of post main sequence stars in M13's core, study the relative radial 
distributions of the different evolved stellar populations in M13 and in two
other dense clusters, and compare M13's stellar luminosity function to a
theoretical luminosity function.  The conclusions of the paper are summarized
in Sec.~4.

%\vfill\eject
\section{THE DATA}

\subsection{Observations}

Two 60$\,$s exposures of M13 were obtained in January~1991 with {\it HST\/}'s
Planetary Camera-I, one through each of the F555W and F785LP filters.
These bandpasses are roughly similar to the Johnson $V$ and $I$ bands,
respectively (Harris et~al.\ 1991).  Each PC image consists of a $2\times2$
mosaic of $800\times800$ CCD images (PC5--PC8) with a pixel scale of about
$0\farcs044$.  A greyscale representation of the F555W mosaic image of M13 is
shown in Figure~1.  The field of view of the CCD mosaic image is
$68^{\prime\prime}\times68^{\prime\prime}$ with $\sim0\farcs5$-wide unusable
strips between adjacent CCD frames.  The telescope was pointed so that the
center of the cluster was imaged on PC6; the pointing was identical for the
F555W and F785LP exposures.  The gain of the PC CCDs is
7.6~electrons~ADU$^{-1}$ and the read noise is about 13~electrons.  The
limited dynamic range of the analog-to-digital converters results in
saturation of the raw CCD frames at the level of 4096~ADU
($3\times10^4~$electrons).  Details of the instrument may be found in
Griffiths (1989), Burrows et~al.\ (1991), and Faber (1992); see Paper~I for
details of the instrumental configuration used for these observations.

\subsection{Data Processing}

The images were processed via the standard Space Telescope Science Institute
pipeline, the primary elements of which were overscan correction, bias
(``zero'') subtraction, analog-to-digital bit correction, and flat fielding
(Lauer 1989).  Bad columns were visually identified and interpolated over.
Bias subtraction and flat fielding lower the saturation level to
$\sim3500$~ADU and cause the saturation level to vary by $\lesssim10$\%
across the image.  At most 5~pixels at the centers of the point spread
function (PSF) cores of the brightest stars in the F555W image
($V\lesssim14.5$) and at most 2~pixels at the centers of the brightest stars
in the F785LP image ($I\lesssim12.2$) were saturated.  Saturated pixels were
masked and subsequently ignored, and only the unsaturated wings of the 
saturated bright stars were used in the PSF building process and in the PSF
fit to derive stellar photometry (see below).

The spherical aberration of the pre-refurbishment {\it HST\/} optics
produces a spatially-varying PSF with a complex shape (Fig.~1): a
sharp core ($\rm FWHM=0\farcs1$) contains $\sim10$\% of the light
while the rest is distributed in a faint, but complex halo extending
out to $r_{\rm PSF}\approx2\farcs5$ (Burrows et~al.\ 1991).  Faint
stars located within the PSF wings of bright stars are measured
reliably only if the model of the PSF accurately accounts for the
light in the wings of the bright star.  Papers~I and II describe a
star-finding, PSF-building, and PSF-fitting procedure based on
{\smcap daophot} and {\smcap daophot~ii} (Stetson 1987, 1992),
designed to extract accurate stellar photometry under crowded
conditions (i.e.,~where there is significant overlap between the PSFs
of stars in the image).

The {\smcap find} routine of the {\smcap daophot} package, a matched
filter convolution technique, was used to find stars on the F555W and F785LP
images.  The {\smcap daophot} detection list for each CCD image was
manually edited to remove spurious detections (e.g.,~the PSF tendrils of
bright stars) and to add stars fainter than the {\smcap find} threshold,
based on an inspection of the residual image, the difference between the
original image and the best fit PSF template (see below).  The reader is
referred to Papers~I and II for details.

The number of bright stars in the PC image of M13 suitable for building a PSF
template is small (15--20)---at least 30~bright, relatively isolated stars
per CCD image are needed to reliably build a quadratic template.  Because of
this, it was deemed best {\it not\/} to follow the full PSF-building process
described in Paper~I.  We instead experimented with a pre-existing set of PSF
templates derived from our earlier 47~Tuc, M15, and M3 images obtained in
March~1991, April~1991, and July~1992, respectively (Papers~I, II, and III,
respectively).  Each set consists of eight quadratically variable templates
(as defined in the {\smcap daophot~ii} package), one for each of CCDs
PC5--PC8 and for each of the F555W and F785LP images.  Each of the three
pre-existing sets of PSF templates was fit to the stars in the M13 detection
list using {\smcap daophot~ii}'s PSF-fitting program {\smcap allstar}.  While
the photometric lists were roughly similar for the three sets of templates,
the M3 PSF templates produced residual images (original minus PSF template)
that were slightly smoother than those obtained with the other two sets of
templates.  In fact, the M13 residual image is comparable in quality to the
residual images of the clusters in our earlier studies in which the PSF
template was derived in an internally consistent fashion (i.e.,~from stellar
images in the dataset).

The eight M3 PSF templates (PC5--PC8, F555W and F785LP) were used to make a
simultaneous fit to all the resolved stars on the corresponding M13 PC CCD
images, yielding positions and instrumental magnitudes in the process (see
Paper~I for details).  An F555W vs $\rm F555W-F785LP$ color--magnitude
diagram (CMD) was obtained for each of the four PC CCDs.  These CMDs were
used for matching the zeropoints of the F555W and F785LP instrumental
magnitude scales between the four CCDs, using as reference points the
brightness of the horizontal branch (HB) and the color of the red giant
branch (RGB) at the level of the HB.  Typical inter-CCD zeropoint adjustments
amounted to $\lesssim\pm0.05$~mag, consistent with the values found in our
earlier studies (Papers~I and II).

\subsection{Completeness and Photometric Accuracy}

The F555W simulations that we carried out for our earlier 47~Tuc study
(Paper~I) can be used to assess the degree of completeness and level of
photometric accuracy in the M13 dataset.  The difference between the apparent
distance moduli of the two clusters (difference between their $V_{\rm HB}$)
is $\Delta(V-M_V)_{\rm obs}\sim+0.9$~mag, with M13 being fainter.  Most of
this difference is attributed to the difference in their distances from the
Sun [$D_{\rm 47\,Tuc}=4.6$~kpc; $D_{\rm M13}=7.2$~kpc;
$\Delta(V-M_V)_0=+0.97$~mag], since neither cluster is appreciably reddened
[$(A_V)_{\rm 47\,Tuc}=0.12$~mag; $(A_V)_{\rm M13}=0.06$~mag].  The shape of
the observed $V$-band stellar luminosity function (LF) of post main sequence
stars is very similar for the two clusters, after one accounts for the
0.9~mag shift between their apparent $V$ magnitude scales.  The {\it HST\/}
PC exposure times, $t_{\rm 47\,Tuc}=26\,$s and $t_{\rm M13}=60\,$s, were
chosen to exactly compensate for this magnitude difference: $2.5\,{\rm
log}(t_{\rm M13}/t_{\rm 47\,Tuc})=0.91$~mag.  Since sky noise is unimportant
in these short exposures, this implies that the ratio of signal to (read)
noise is practically identical for two stars of a given {\it absolute\/} $V$
magnitude in the two clusters.

As described in Paper~I, we have carried out simulations in the F555W band
with a variety of projected stellar densities: 2000, 1000, and
500~simulations (these numbers refer to the approximate number of stars
brighter than the main sequence turnoff within the $\sim0.3$~arcmin$^2$ field
of view of a single PC CCD).  The 2000/F555W~simulation mimics the stellar
surface density in the core of 47~Tuc; the projected stellar density in M13's
core is about an order of magnitude lower.  The 1000/F555W and
500/F555W~simulations indicate that the M13 F555W dataset is $>90$\% complete
for stars with $\rm F555W<19$ (shifting the magnitude scale in Fig.~10 of
Paper~I by +0.9~mag).  This is consistent with M13's observed stellar LF in
the F555W band, which continues to rise to $\rm F555W\sim19$, but
drops sharply due to incompleteness beyond $\rm F555W=20$ (Fig.~2).  It is
evident from Figure~2 that the magnitude at which incompleteness sets in is
independent of radial distance from the cluster center; this is due to the
fact that M13's core covers a large fraction of the area of the four-CCD PC
mosaic so that the stellar surface density (or the degree of crowding) only
varies by a factor of~4 across the PC image.

The rms photometric uncertainty in the F555W band is expected to be
$1\sigma\lesssim0.1$~mag for stars with $\rm F555W<19$, with a somewhat
non-Gaussian error distribution (see Figs.~5 and 6 of Paper~I).  This is in
rough agreement with estimates of the photometric error derived from the
observed width of M13's RGB and HB, assuming these have zero intrinsic spread
(Fig.~3).  Systematic photometric error due to PSF template mismatch in M13
is comparable to that in our earlier cluster studies and simulations judging
from the smoothness of the M13 residual images and from the width of its
bright RGB (for which the scatter is dominated by systematic error).

The intrinsic shape of the RGB---specifically, the fact that its color gets
redder towards the bright end---implies that the brightness ratio between the
brightest (bright RGB) stars and faintest (subgiant/turnoff) stars is greater
in the $I$ band than in the $V$ band.  Detection and photometry of faint
stars are affected by the proximity of brighter stars; these crowding effects
are somewhat worse in $I$ than in $V$.  Results from the earlier 47~Tuc
1000/F785LP~simulation indicate that incompleteness in the matched sample of
F555W and F785LP detections sets in about 1~mag brighter than in the sample
of `F555W-only' detections (Fig.~10 of Paper~I).  Note that M13's bright RGB
does not curve over to the red as much as 47~Tuc's since its metallicity is
lower than that of 47~Tuc; consequently, the results from the 47~Tuc F785LP
simulation may not be applicable to M13 in detail.  Figure~2 shows the LFs of
the F555W-only sample and the matched F555W and F785LP sample in the central
part of M13.  The LF of the matched sample turns over due to incompleteness
at $\rm F555W\sim18.5$, nearly 1~mag brighter than in the F555W-only LF.  We
return to a discussion of the shape of M13's LF in Sec.~3.5.

%\vfill\eject
\section{STELLAR POPULATIONS}

\subsection{V and I Band Stellar Photometry}

The PSF-fitting procedure described in Sec.~2.2 yields a list of positions
and F555W brightnesses for 6436~stars in M13 down to $V\sim21$ within the
1.25~arcmin$^2$ area of the PC mosaic.  The dataset includes stars out to
$r<66^{\prime\prime}$, but only annuli with $r\lesssim14^{\prime\prime}$ are
completely contained within the PC image.  Of the 6436~F555W
detections, 2877 have a corresponding star within $\Delta{r}<0\farcs14$ on
the F785LP image.  This list of 2877~matched F555W and F785LP detections
appears to be complete to about $V\sim18.3$, and hardly any stars with $V>20$
are detected on the F785LP image (Sec.~2.3, Fig.~2).

The WF/PC-I instrumental calibration constants (Faber 1992) are used to
convert the counts (in ADU) reported by the PSF-fitting program into
instrumental F555W and F785LP magnitudes.  These were then converted to
Johnson $V$ and $I$ magnitudes, respectively, using the color transformation
equations obtained by Harris et~al.\ (1991).  The zeropoints of the F555W and
F785LP magnitude systems are roughly equal to those of Johnson $V$ and $I$,
respectively, and there are small but non-zero color terms.  Thus, it is only
possible to derive magnitudes on the Johnson system for those stars detected
in {\it both\/} F555W and F785LP.

Figure~3 is a $V$ vs $V-I$ CMD of all stars matched in F555W and F785LP in
our M13 sample.  The division into different stellar types is only
approximate and is based on the gross features of the distribution of stars
in the CMD.  The data extend below the main sequence turnoff ($V\sim18.3$,
$V-I\sim0.4$), although they are increasingly incomplete below it.  The RGB
gets only slightly redder towards its bright end, and the HB droops by almost
3~mag in $V$ and extends quite far to the blue ($V-I\sim-0.5$); these
characteristics of M13's CMD are more or less in keeping with its relative
low metallicity: $\rm [Fe/H]=-1.65$ (Djorgovski 1993).  A sparse group of
asymptotic giant branch (AGB) stars is visible at $V\sim14$ and
$V-I\lesssim0.8$.  A small number of blue straggler stars (BSS) lie in the
region slightly brighter and bluer than the main sequence turnoff:
$V\lesssim18$ and $V-I\lesssim0.4$.  The large scatter at the faint end
($V\gtrsim18.5$) is mostly a result of photometric error.

We have compared the $V$ vs $V-I$ CMD of M13's core (Fig.~3) with
ground-based photometry in the $B$ and $V$ bands (Guarnieri et~al.\ 1993;
Sandage 1970) and in the $B$ and $I$ bands (Stetson 1996) of stars at a
variety of radii in the cluster.  While a star-by-star comparison is not
possible, a comparison of several fiducial points in the CMD (e.g.~RGB color
at various levels, HB magnitude and color) indicates that the $V$ and $I$
magnitude zeropoints we have adopted agree with ground-based photometry to
within $\sim\pm0.1$~mag.  For the most part however, our study relies on
relative photometry of stars and not on absolute photometry.

The positions and brightnesses of a subset of 25~bright ($V\lesssim13.5$),
relatively isolated stars in the core of M13 are listed in Table~1.  All
positions are on the equinox~J2000 coordinate system, and are measured
relative to the reference star~U (ID\#5540) whose coordinates are:

\begin{equation}
\rm\alpha_{J2000}(U)=16^h41^m41\fss61;~~~
\delta_{J2000}(U)=+36^\circ27^\prime42\farcs06
\end{equation}

\noindent
The coordinates of star~U (accurate to about $0\farcs25$) and the position
angle of the WF/PC-I images ($38\fdeg46\pm0\fdeg5$; see Fig.~1) are
based on the {\it HST\/} Guide Star Catalog.  We adopt Gould \& Yanny's
(1994) prescription for the pixel scale ($\sim0\farcs044$~pixel$^{-1}$) and
inter-CCD offsets and rotations (Fig.~1) of the PC.  The bright stars A--Y
listed in Table~1 are marked on the greyscale image of the cluster in
Figure~1 and serve as a ``finding chart'' to facilitate matching our dataset
to others.  The full version of Table~1 is presented in the AAS CD-ROM
Series, volume~X, 199X.  It contains all 6436~stars detected in the F555W
band (with ID\#s sorted by right ascension), including the 2877~stars matched
in the F555W and F785LP lists for which F555W, F785LP, $V$, and $I$
magnitudes are provided.

\subsection{Blue Stragglers}

The $V$ vs $V-I$ CMD of M13 presented in Figure~3 shows a sparse collection
of BSS candidates in the region between the main sequence turnoff and the
drooping tail of the extended blue HB.  About 15~of these BSS candidates lie
roughly where the continuation of the upper main sequence would be, if it
were to be extrapolated beyond (i.e.,~brighter and bluer than) the turnoff.
These 15~stars are reasonably distinct from the HB and turnoff stars in the
CMD and are to be regarded as ``likely'' candidates.  An additional 10~BSS
candidates are indicated by the dotted lines in Figure~3.  These 10~stars lie
close to M13's HB or subgiant branch; some of them may be HB or subgiant
stars that have been scattered away from their fiducial locations in the CMD
due to measurement error.  Thus, these additional 10~BSS are to be considered
``possible'' candidates.  Unless otherwise mentioned, we use the term BSS
candidates to refer to all 25~stars.

The actual number of BSSs in the region of M13 covered by our PC image is
likely to be in the range 15--25.  As discussed in Sec.~2.3, the list of $V$
and $I$ detections should be $>90\%$ complete down to the main sequence
turnoff which makes it unlikely that a BSS would go undetected.  Our BSS
selection criterion (Fig.~3), while consistent with the criteria used in our
earlier studies (Papers~I--III) and in previous ground-based studies
(cf.~Bolte et~al.\ 1993; Stetson 1994), is somewhat arbitrary.  In
particular, it excludes the less extreme BSSs---those with $V-I$ colors
within 0.3~mag and/or $V$ magnitude within 0.5~mag of the turnoff.  It is
likely that we compensate (possibly overcompensate) for the exclusion of less
extreme BSSs from our sample by the including a few ``spurious''
objects---i.e.,~turnoff, subgiant, and HB stars that have scattered into the
BSS region due to photometric error.  A comparative study of 47~Tuc's dense
core shows that, of the complete sample of BSSs detected in the ultraviolet
study of Paresce et~al.\ (1991), only about half are extreme enough to appear
as definite BSSs in a $V$ vs $V-I$ CMD, clearly separated from the turnoff
and subgiant branches (Paper~I).

The specific frequency of BSS in M13 is calculated according to the
definition of Bolte et~al.\ (1993): $F_{\rm BSS}=N({\rm BSS})/N(V<V_{\rm
HB}+2)$.  Although M13's HB is not truly horizontal in the CMD (Fig.~3), it
has a well defined brightness of $V\approx15$ at the location of the RR~Lyrae
instability region before drooping to fainter $V$ magnitudes at the blue end.
There are 343~subgiants in the M13 PC image that fulfil the criterion for the
normalizing population.  The specific frequency is $F_{\rm BSS}=0.04$ if we
count only the 15~definite BSS candidates in M13's core, and $F_{\rm
BSS}=0.07$ if we count all 25~candidates.  The $1\sigma$ Poisson error in the
specific frequency is $\delta{F}_{\rm BSS}=0.01$, but systematic errors in
the BSS sample selection are likely to be somewhat larger.

Table~2 compares M13's BSS frequency to that of a few other globular clusters
(listed in order of decreasing stellar density) that have been studied
recently.  As pointed out by Sosin \& King (1995), it is interesting that
the BSS frequency in this representative set of clusters spans only a range
of a factor of~three, with a mean value of $F_{\rm BSS}\sim0.11$, despite the
$\ga10^4$~range in stellar density among the clusters.  M13's BSS frequency
is also within this relatively narrow range.

\subsection{Radial Distribution of Stellar Populations}

In this section, we compare the radial density distributions of the various
types of evolved stars found in M13's core.  The cluster core fills most of
the field of view of the PC mosaic, so our sample spans only a small range of
stellar space densities (roughly a factor of~8).  We divide the sample of
stars into four (projected) radial bins:
(1)~$r<15\farcs0$,
(2)~$15\farcs0\leq{r}<25\farcs8$,
(3)~$25\farcs8\leq{r}<37\farcs6$, and
(4)~$37\farcs6\leq{r}<65\farcs6$,
around the cluster center.  We define the cluster center to be the centroid
of the light distribution as measured on a ground-based image of M13:

\begin{equation}
\rm\alpha_{J2000}(0)=16^h41^m41\fss2;~~~
\delta_{J2000}(0)=+36^\circ27^\prime35^{\prime\prime}
\end{equation}

\noindent
These coordinates are within $4^{\prime\prime}$ of the cluster centroid
coordinates quoted by Djorgovski (1993), well within the measurement errors.
The limiting radii of the four~radial bins have been chosen so that each bin
contains the same number of stars with $V<18$.  This relatively bright
subsample of stars, dominated in number by subgiants and faint RGB stars, is
complete at all radii (Sec.~2.3) and serves as a convenient normalizing
population in studying the radial distribution of various stellar types.  The
first three radial bins are entirely within the core of M13 ($r_{\rm
core}=38^{\prime\prime}$).  The rectangular field of view of the PC includes
only a fraction of the annulus for bins 2, 3, and 4 (56\%, 42\%, and 18\%,
respectively).

Figure~4 is a composite of the CMDs in each of the four radial bins.  There
are no striking differences in the mix of stellar types between annuli.  The
innermost bin appears to be somewhat deficient in HB stars (quantified
below).  The turnoff portion of the CMD has slightly less scatter and a
higher number of stars as one moves radially outwards, a result of improved
photometric accuracy and increased degree of completeness in the sparser
parts of the cluster.

Table~3 lists the number of stars of each type in the four radial bins. The
division of the post main sequence stars in M13 into various stellar types is
defined by the dashed lines in Figure~3.  The total number of stars increases
towards larger radii due to the increased probability of detecting faint
turnoff stars and main sequence stars ($V\sim20$).  Faint RGB stars and
subgiants make up the bulk of the normalizing population used to define the
limiting radii, so it is not surprising that their numbers are roughly
constant across the four bins.  There is no radial gradient in the fraction
of bright RGB or BSS stars either.  Poisson errors are very large for the BSS
population; small changes in the BSS fraction as a function of radius
(smaller than a factor of 2--3) would be undetectable in our sample of
25~candidates.  The HB stars show a marginal central depletion.  Their
expected number per bin is $21\pm4.6$ ($1\sigma$).  The HB population of the
innermost radial bin appears to be about 55\% of the expected number, and is
low by $-1.7\sigma$.
% The central deficit seems to be a little more pronounced for stars in the
% drooping blue extension of the HB ($V\gtrsim16$, $V-I\lesssim-0.25$) than
% for the redder HB stars; the small number of HB stars, however, makes this
% conclusion uncertain.

The cumulative radial distributions of different types of stars in the core
of M13 are plotted in Figure~5.  A two-sided Kolmogorov--Smirnov test shows
that the radial distributions of the faint RGB/subgiant, bright RGB/AGB, and
BSS populations are consistent with one another.  The HB stars show a
marginal central depletion relative to the faint RGB/subgiant stars (82\%
significance level or $\sim1.5\sigma$).  The radial distribution of faint
RGB/subgiant stars in M13 is well fit by a King profile (King 1962) of core
radius $r_{\rm core}=38^{\prime\prime}$ (1.3~pc) as indicated by the thin
solid curve in Figure~5.  The 90\% confidence limit on the core radius of
M13's RGB stars is $\pm6^{\prime\prime}$.  The HB stars, on the other hand,
have a core radius of $r_{\rm core,\,\>HB}\gtrsim60^{\prime\prime}$.  It is
perhaps not surprising that Trager et~al.\ (1993) found a best fit core
radius of $r_{\rm core}=53^{\prime\prime}$ for the overall surface brightness
distribution (intermediate between the RGB and HB core radii), since RGB
stars and HB stars together account for most of the visual light in M13.

\subsection{Radial Population Gradients in Three~King~Model Clusters:
M13,~M3,~and~47~Tuc}

Subtle (but real) radial gradients in the stellar populations of
globular clusters may be masked by large statistical errors in the
measurement of the stellar density that result from having a limited
number of stars in the core of a single cluster.  To reduce the
Poisson errors associated with the measurement of population
gradients, we combine samples of post main sequence stars from five
dense clusters: M13, M3, 47~Tuc, M30, and M15.  The cumulative
distributions of the subsamples of faint RGB/subgiant
and HB stars in this combined sample are plotted as a function of
the projected $r/r_{\rm half~light}$ in the upper panel of Figure~6.
The lower panel of Figure~6 shows the cumulative radial distribution vs
$r/r_{\rm core}$ for the combined faint RGB/subgiant and HB
samples from the three King model clusters only: M13, M3, and 47~Tuc.

The following values of $r_{\rm half~light}$ ($r_{\rm core}$) have
been adopted: $93^{\prime\prime}$ ($38^{\prime\prime}$, Sec.~3.3) for M13,
$68^{\prime\prime}$ ($28^{\prime\prime}$, Paper~III) for M3,
$174^{\prime\prime}$ ($23^{\prime\prime}$, Paper~I) for 47~Tuc,
$62^{\prime\prime}$ for M30, and $60^{\prime\prime}$ for M15.
Half light radii are taken from Trager et~al.\ (1993); M30 and M15 do not
have well defined core radii.  The data extend out to
$66^{\prime\prime}$ ($0.7\,r_{\rm half~light}$, $1.7\,r_{\rm core}$) in M13, 
$65^{\prime\prime}$ or $1.0\,r_{\rm half~light}$ ($2.3\,r_{\rm core}$) in M3, 
$61^{\prime\prime}$ or $0.4\,r_{\rm half~light}$($2.7\,r_{\rm core}$) in
47~Tuc, $130^{\prime\prime}$ or $2.1\,r_{\rm half~light}$ in M30, 
and $130^{\prime\prime}$ or $2.2\,r_{\rm half~light}$ in M15.

There are 2648~faint RGB/subgiant stars in the five-cluster sample
(248~from M13, 375~from M3, 561~from 47~Tuc, 290~from M30, and 1174~from M15)
and 1184 in the three-cluster sample; the distribution of faint RGB/subgiant
stars among the clusters
indicates the relative weight given to each cluster in the combined
cumulative distribution.  Note that the data for the five clusters extend
out to different values of ${r/r_{\rm half~light}}$ and ${r/r_{\rm
core}}$.  Thus, it is necessary to ensure that the ratio of faint
RGB/subgiant to HB (or BSS) stars is
the same for each cluster to avoid radially biasing the population ratio.
We adjust the size of the faint RGB/subgiant sample in each cluster (by
changing the faint end limiting magnitude) so that the ratio of HB (or BSS)
to faint RGB/subgiant stars is the same in each cluster.

Two-sided Kolmogorov--Smirnov tests were performed on the cumulative radial
distribution functions of the combined three- and five-cluster samples for
the BSS vs faint RGB/subgiant and HB vs faint RGB/subgiant stellar
populations.  The BSSs are marginally centrally concentrated relative to the
normalizing population (85\% confidence level) in the three-cluster sample.
The central concentration of BSSs in the combined three-cluster sample is of
comparable significance to that in M3 (86\% confidence---Paper~III) and is
{\it less\/} significant than in 47~Tuc (90\%--95\% confidence---Paper~I);
the nature of the BSS concentration is sufficiently different in M3 and
47~Tuc that the effect is not reinforced in the combined sample.  In the
five-cluster sample, the BSSs are significantly more centrally concentrated
than the normalizing faint RGB/subgiant population: the probability that the
two populations are drawn from the same radial distribution is only
$10^{-5}$; this is largely driven by the numerous, strongly centrally
concentrated BSSs in M30.

The HB stars in the combined sample show some evidence for central
depletion relative to the faint RGB/subgiant stars.  The depletion
is significant at the 85\% confidence level in the three-cluster
sample ($\sim1.5\sigma$), and at the 97\% level in the five cluster-sample
($\sim2\sigma$).  Is interesting to
note that each of the the five dense globular clusters that we have studied
in our {\it HST\/} program so far displays a weak (and only marginally
significant in each individual cluster) central HB deficiency: M13 (Fig.~6),
M3 (Fig.~12 of Paper~III; Bolte et~al.\ 1993), 47~Tuc (Fig.~18 of Paper~I),
M30 (Fig.~5 of Yanny et~al.\ 1994b), and M15 (Fig.~18 of Paper~II).  As
noted by Laget et~al.\ (1992), studies of M14, M15, and NGC~6397
suggest that red HB stars (those redder than the RR~Lyrae gap) may be
more centrally concentrated than blue HB stars; our samples indicate
no obvious difference between the degree of depletion of red HB stars
(47~Tuc) and blue HB stars (M30, M13) in cluster cores.  Furthermore,
a WFPC2 study of M15 shows that the red and blue HB stars in this
cluster have similar radial density distributions (Guhathakurta
et~al.\ 1996).

The cause of the central HB depletion in dense globular cluster cores remains
a mystery.  It is unlikely to be due to mass segregation between HB stars and
the more massive subgiants because: (1)~the dynamical time scale for mass
segregation in the cores of these dense clusters, $t_{\rm dyn}\approx10^9~$yr
(Djorgovski 1993), is larger than the time spent by a star in the HB phase,
$t_{\rm HB}\approx10^8~$yr (Lee \& Demarque 1990); and (2)~standard stellar
evolution theory (Lee et~al.\ 1990) predicts only a small amount of mass loss
in the RGB phase and hence only a small difference in mass between an HB star
($\gtrsim0.7\,M_\odot$) and a subgiant ($0.8\,M_\odot$).  

The central HB depletion is not an artifact of a higher rate of
misidentification of HB stars near the cluster center than in the outer
parts.  Firstly, for relatively bright stars such as HB stars, the
photometric errors are fairly small ($1\sigma\sim0.1$~mag) and do not vary
appreciably with distance from the cluster center (especially in the case of
M13 where the stellar density does not vary drastically over the area of the
PC image).  Secondly, the section used to define HB stars in the CMD
(cf.~Fig.~3) is large enough that HB stars are unlikely to scatter out of the
region because of photometric error.

\subsection{Luminosity Function}

Figure~7 shows the LF of evolved stars in the PC image of M13 (open bold
symbols).  The LF is plotted versus stellar F555W magnitude, and not the
transformed Johnson $V$ magnitude which requires both F555W and F785LP
measurements.  As discussed in Sec.~2.3, the sample of F555W-only detections
from which the LF is derived is more complete at the faint end than the
sample of ``matched'' F555W and F785LP detections.  Incompleteness sets in
fainter than $\rm F555W\sim19$ in the F555W-only sample; at brighter
magnitudes, the LF shape appears to be independent of radial distance from
the center of M13 (Fig.~2).  We therefore combine data from the entire PC
image to minimize Poisson error ($1\sigma$ error bars are shown in Fig.~7).
Even though detection in the F555W band is the only criterion used to select
stars for the LF, in practice, the brightest stars are detected in both F555W
and F785LP bands.  For reasons explained below, we use a color cut
($V-I<0.4$) to exclude HB stars and to restrict the bright part of the LF
($\rm F555W<17.3$) to only RGB and AGB stars.

The dashed line in Figure~7 shows a theoretical model for the LF of an old
(16~Gyr), metal-poor ($\rm [Fe/H]=-1.66$) stellar population (Bergbusch \&
Vandenberg 1992).  These parameters have been chosen to match the age
($14.6\pm3.0$~Gyr---Carney et~al.\ 1992) and metallicity ($\rm
[Fe/H]=-1.65$---Djorgovski 1993) of M13.  The absolute visual magnitude scale
in the model LF has been converted to the apparent F555W magnitude scale
using a distance modulus of $(m-M)_0=14.29$~mag and a Galactic extinction of
$A_V=0.06$~mag for M13 (Djorgovski 1993).  The model LF has been normalized
to match the observed number stars at $\rm F555W=19$, (conservatively)
assuming that the M13 sample is only 90\% complete for stars of this apparent
brightness (see Sec.~2.3).  The Bergbusch--Vandenberg model does not include
the HB phase of stellar evolution; hence we have excluded HB stars from M13's
LF before comparing it to the model LF.  Fainter than $\rm F555W=19$, the
observed LF falls below the model LF due to the effect of incompleteness, as
predicted by our image simulations.

A comparison between M13's LF and the Bergbusch--Vandenberg model LF (Fig.~7)
shows that the model underpredicts the number of bright RGB stars relative to
the subgiants.  Specifically, there are twice as many bright RGB stars in the
range $V=12.5$--15 in M13 than predicted by the model, a $6.4\sigma$
difference (taking only Poisson error into account).  Note, the M13 LF
includes both RGB and AGB stars at its bright end (since the PC photometry is
not accurate enough to allow a clear distinction between these two classes)
whereas the Bergbusch--Vandenberg model LF includes only RGB stars.  In order
to resolve the difference between the model and data, roughly half the
bright stars in M13 ($V\lesssim15$) would have to be AGB stars.  This seems
unlikely since AGB stars typically make up a small fraction of the bright end
of the LF in globular clusters: $N({\rm AGB})/N({\rm bright~RGB}) \approx
N({\rm AGB})/N({\rm HB})\approx0.15$ (Renzini \& Fusi Pecci 1988; Buzzoni
et~al.\ 1983; Buonanno et~al.\ 1985).  Furthermore, we have obtained a
ground-based CMD of stars in M13 with sufficient photometric accuracy to
distinguish AGB stars from bright RGB stars, and this indicates a comparably
small AGB fraction at the bright end of M13's LF.

The discrepancy in the number of bright RGB stars between M13 and the model
is not a result of incorrect normalization of the model; the model is in
excellent agreement with the data for the faint RGB, subgiant, and turnoff
stars ($\rm F555W=15$--18.5).  The horizontal offset of the model LF (along
the F555W magnitude scale), derived from the quoted M13 distance modulus and
line-of-sight visual extinction, is also in good agreement with the data
judging from the sharp upturn in both LFs at the base of the giant branch at
$\rm F555W\approx17$.  Moreover, varying the metallicity of the model over
the range, $\rm-2\lesssim[Fe/H]\lesssim-1$, and its age over the range,
$\rm8~Gyr<{\it t}<18~Gyr$, does not make a significant difference to the
quality of the fit.

Discrepancies between the relative numbers of RGB stars and turnoff stars
have also been noted in the clusters M92 (Stetson 1991) and M30 (Bolte 1994).
Unlike the case of M13, where there is only a discrepancy for the brightest
RGB stars, these other two clusters display an excess of faint as well as
bright RGB stars relative to turnoff stars.  Larson et~al.\ (1995) suggest
that a modification of the canonical stellar model to include the effects of
internal rotation might boost the lifetimes of faint RGB stars and subgiants
enough to resolve the discrepancy for such stars; their models, however, do
not examine the effects of rotation in bright RGB stars (those at or above
the HB level) and therefore do not address the problem of M13's LF directly.
Interestingly, a recent study of M5 shows good agreement between the observed
LF and that predicted by the canonical Bergbusch--Vandenberg model (Sandquist
et~al.\ 1996).  A systematic study of the LFs of evolved stars in globular
clusters is needed to determine what fraction of clusters display this RGB
excess and what global property of the cluster might be correlated with the
excess.  A thorough investigation of model parameter space is currently in
progress (Vandenberg 1996, private communication).  For now, the question of
what causes the RGB discrepancy remains unanswered.

\section{SUMMARY}

\noindent
The principal results obtained in this paper are summarized below:

\begin{itemize}

\item[{\bf 1.}]{{\it Hubble Space Telescope\/} (pre-refurbishment) Planetary 
Camera-I images in the F555W ($V$) and F785LP ($I$) bands of the central
$68^{\prime\prime}\times68^{\prime\prime}$ region of the dense globular 
cluster M13 (NGC~6205) have been analyzed.  A table is presented containing
relative astrometry and photometry in the F555W, F785LP, $V$, and $I$ bands
of 2877~post main sequence stars within $r<66^{\prime\prime}$ of the cluster
center, and F555W photometry of an additional 3559~faint stars
($V\approx18$--21).}

\item[{\bf 2.}]{Fifteen blue straggler candidates and 10 other possible BSS
candidates are identified in the core of M13 on the basis of a $V$ vs $V-I$
color--magnitude diagram.  The specific frequency of BSS in M13, $F_{\rm
BSS}\equiv{N}({\rm BSS})/N(V<V_{\rm HB}+2)=0.04$--0.07, is comparable to the
value measured in other dense clusters.  The radial distribution of BSSs is
consistent, within Poisson error, with that of other stellar types.}

\item[{\bf 3.}]{The radial surface density profiles of the bright red giants,
faint red giants, subgiants, and turnoff stars are consistent with one
another.  These stars are well fit by a King profile of core radius $r_{\rm
core}=38^{\prime\prime}\pm6^{\prime\prime}$ (1.3~pc).  Horizontal branch
stars appear to be centrally depleted relative to the giants and subgiants
(only a $\gtrsim1.5\sigma$ effect), and their distribution is better described
by an $r_{\rm core,\,\>HB}\gtrsim60^{\prime\prime}$ King profile.} 

\item[{\bf 4.}]{There is a hint of a slight central HB deficiency in each of
the five dense globular clusters we have studied to date: M13, M3, 47~Tuc,
M30, and M15; the significance of this effect in each cluster, however, is
marginal.  Samples of various stellar types from these clusters are combined
to derive the cumulative radial distribution as a function of $r/_{\rm
half~light}$; the three King model clusters, M13, M3 and 47~Tuc are combined
in terms of $r/_{\rm core}$.  The HB stars appear to be less centrally
concentrated than the giants (97\% and 85\% significance in the five- and
three-cluster samples, respectively).  The BSS stars in the combined
five-cluster sample are more centrally concentrated than the giants
($>5\sigma$ effect).}

\item[{\bf 5.}]{The stellar luminosity function of M13 in the F555W ($V$)
band is compared to a suitably normalized, standard model LF for an old,
metal-poor population (Bergbusch \& Vandenberg 1992).  The model is a good
match to the shape of the observed LF for stars in the brightness range
$V\sim15$--19 (the M13 sample is incomplete at fainter magnitudes).  However,
the relative number of bright red giants in M13 ($V\lesssim15$) is a factor
of two higher than predicted by the model.}

\end{itemize}

\bigskip
\bigskip
We would like to thank John Faulkner, Eric Sandquist, and Dennis Zaritsky for
useful discussions and the referee, George Djorgovski, for helpful tips on
combining cluster samples.  We are grateful to Zodiac Webster for allowing us
to include unpublished work on M30 in Sec.~3.4.  P.G.\ would like
to thank the Institute for Advanced Study for its generous
hospitality.  This research was supported in part by NASA through
Grant No.\ NAG5-1618 and Grant No.\ HF-1033.01-93B from the Space
Telescope Science Institute, which is operated by the Association of
Universities for Research in Astronomy, Inc., under NASA Contract No.\
NAS5-26555.

\section*{APPENDIX: TABLE OF STARS}

The full version of Table~1 contains photometry in the F555W band (similar,
but not identical, to $V$) and relative astrometry for 6436~evolved stars
down to $V\sim21$ in M13's core region.  In addition, F785LP, $V$, and $I$
photometry is presented for 2877~of these stars.

A subset of Table~1, containing the 25~brightest stars in the core of M13,
appears in this paper; the full version is published in the AAS CD-ROM
Series, volume~X, 199X.  Computer-readable copies of the full table may be
obtained via anonymous ftp as follows:

\begin{itemize}
\item[$\bullet$]{{\tt ftp eku.sns.ias.edu} ~~(login as: {\tt anonymous})}
\item[$\bullet$]{\tt cd pub/GLOBULAR\_CLUSTERS/m13}
\item[$\bullet$]{\tt get tbl.cdrom}
\end{itemize}

\noindent
or by contacting P.G.\ or R.L.C..

Details of the photometric procedure and estimates of the completeness and
photometric accuracy may be found in Sec.~2 and references therein.  The
reader is referred to Sec.~3.1 for a brief description of the astrometric
convention and the parameters used in astrometric solution (see also Fig.~1).
We recommend that the astrometry provided in Table~1 be matched empirically
against other datasets, to correct for possible systematic errors in the
scale, translation, and rotation of the coordinate system we have adopted.

%\vfill\eject

\section*{References}

\reference Baum, W.~A., Hiltner, W.~A., Johnson, H.~L., \& Sandage, A.~R.
1959, ApJ, 130, 749

\reference Bergbusch, P.~A., \& Vandenberg, D.~A. 1992, ApJS, 81, 163

\reference Bolte, M. 1994, ApJ, 431, 223

%\reference Bolte, M. 1989, ApJ, 341, 168

\reference Bolte, M., Hesser, J.~E., \& Stetson, P.~B. 1993, ApJ, 408, L89

\reference Buonanno, R., Corsi, C.~E., \& Fusi Pecci, F. 1985, A\&A, 145, 97

\reference Burrows, C.~J., Holtzman, J.~A., Faber, S.~M., Bely, P.~Y., Hasan,
H., Lynds, C.~R., \& Schroeder, D. 1991, ApJ, 369, L21

\reference Buzzoni, A., Fusi Pecci, F., Buonanno, R., \& Corsi, C.~E. 1983,
A\&A, 123, 94

\reference Carney, B.~W., Storm, J., \& Jones, R.~V. 1992, ApJ, 386, 663

\reference Cudworth, K.~M., \& Monet, D.~G. 1979, AJ, 84, 774

\reference Djorgovski, S.~G. 1993, in Structure and Dynamics of Globular
Clusters, edited by S.~G.~Djorgovski and G.~Meylan (ASP Conf.\ Series,
No.~50), p.~373

\reference Faber, S.~M. (editor) 1992, Final Orbital/Science Verification
Report, Space Telescope Science Institute Pulication

\reference Gould, A., \& Yanny, B. 1994, PASP, 106, 101

\reference Griffiths, R. 1989, Wide Field and Planetary Camera Instrument
Handbook, Space Telescope Science Institute Pulication

\reference Guarnieri, M.~D., Bragaglia, A., \& Fusi Pecci, F. 1993, A\&AS,
102, 397

\reference Guhathakurta, P., Yanny, B., Bahcall, J.~N., \& Schneider, D.~P.
1994, AJ, 108, 1786 (Paper~III)

\reference Guhathakurta, P., Yanny, B., Schneider, D.~P., \& Bahcall, J.~N.
1992, AJ, 104, 1790 (Paper~I)

\reference Guhathakurta, P., Yanny, B., Schneider, D.~P., \& Bahcall, J.~N.
1996, AJ (in preparation)

\reference Harris, H.~C., Baum, W.~A., Hunter, D.~A., \& Kreidl, T.~J. 1991,
AJ, 101, 677

\reference King, I.~R. 1962, AJ, 67, 471

\reference Laget, M., Burgarella, D., Milliard, B., \& Donas, J. 1992, A\&A,
259, 510

\reference Larson, A.~M., Vandenberg, D.~A., \& Depropris, R. 1995, BAAS, 27,
1431

\reference Lauer, T.~R. 1989, PASP, 101, 445

\reference Lee, Y.-W., \& Demarque, P. 1990, ApJS, 73, 709

\reference Lee, Y.-W., Demarque, P., \& Zinn, R. 1990, ApJ, 350, 155

\reference Lupton, R.~H., \& Gunn, J.~E. 1986, 91, 317

\reference Lupton, R.~H., Gunn, J.~E., \& Griffin, R.~F. 1987, AJ, 93, 1114

\reference Nemec, J.~M., \& Cohen, J.~G. 1989, ApJ, 336, 780

\reference Paez, E., Straniero, O., \& Martinez Roger, C. 1990, A\&AS, 84, 481

\reference Paresce, F., et~al. 1991, Nature, 352, 297

\reference Pryor, C., \& Meylan, G. 1993, in Structure and Dynamics of
Globular Clusters, edited by S.~G.~Djorgovski and G.~Meylan (ASP Conf.\
Series, No.~50), p.~357

\reference Renzini, A., \& Fusi Pecci, F. 1988, ARA\&A, 26, 199

\reference Sandage, A.~R. 1970, ApJ, 162, 841

\reference Sandquist, E.~L. 1996, private communication

\reference Sandquist, E.~L., Bolte, M., \& Stetson, P.~B. 1996, ApJ (in
press)
%
% in Formation of the Galactic Halo, edited by H.~Morrison and A.~Sarajedini
% (ASP Conf.\ Series, No.~92), p.~293

\reference Savedoff, M.~P. 1956, AJ, 61, 254

\reference Sosin, C., \& King, I.~R. 1995, AJ, 109, 639

\reference Stetson, P.~B. 1987, PASP, 99, 191

\reference Stetson, P.~B. 1991, in Formation and Evolution of Star Clusters,
edited by K.~Janes (ASP Conf.\ Series, No.~13), p.~88            

\reference Stetson, P.~B. 1992, in Astronomical Data Analysis Software,
edited by D.~M.~Worrall, C.~Biemesderfer, and J.~Barnes (ASP Conf.\ Series,
No.~25), p.~297

\reference Stetson, P.~B. 1994, PASP, 106, 250

\reference Stetson, P.~B. 1996, preprint
% XXX

\reference Trager, S.~C., Djorgovski, S., \& King, I.~R. 1993, in Structure
and Dynamics of Globular Clusters, edited by S.~G.~Djorgovski and G.~Meylan
(ASP Conf.\ Series, No.~50), p.~347

\reference Yanny, B., Guhathakurta, P., Bahcall, J.~N., \& Schneider, D.~P.
1994a, AJ, 107, 1745 (Paper~II)

\reference Yanny, B., Guhathakurta, P., Schneider, D.~P., \& Bahcall, J.~N.
1994b, ApJ, 435, L39

\reference Zinn 1986, in Stellar Populations, edited by C.~Norman,
A.~Renzini, and M.~Tosi (Cambridge University Press), p.~73

%\end{document}
%\vfill\eject
%\setcounter{page}{26}

\begin{table}
\begin{tabular}{lllllllllll}
\multicolumn{11}{c}{\smcap Table~1.} \\
\multicolumn{11}{c}{Selected Bright Stars in the Core of M13} \\
\hline
\hline

\multicolumn{1}{c} {ID\#$^a$} &
\multicolumn{1}{c} {ID$^b$} &
\multicolumn{2}{c} {Offset in [$^{\prime\prime}$] from U} &
\multicolumn{1}{c} {CCD} &
\multicolumn{1}{c} {$r$} &
\multicolumn{2}{c} {Instr.\ mags.} &
\multicolumn{3}{c} {Johnson mags.} \\

\cline{3-4}
\cline{7-8}
\cline{9-11}

\multicolumn{1}{c} {} &
\multicolumn{1}{c} {} &
\multicolumn{1}{c} {$\Delta\alpha_{\rm J2000}$} &
\multicolumn{1}{c} {$\Delta\delta_{\rm J2000}$} &
\multicolumn{1}{c} {[PC\#]} &
\multicolumn{1}{c} {[$^{\prime\prime}$]} &
\multicolumn{1}{c} {F555W} &
\multicolumn{1}{c} {F785LP} &
\multicolumn{1}{c} {$V$} &
\multicolumn{1}{c} {$I$} &
\multicolumn{1}{c} {$V-I$} \\

\hline

\hfill 2289 & ~A & \hfill $-$28.84 & \hfill $-$18.55~~ & ~~~7 & \hfill 26.9 & ~12.20 & ~~10.66 & 12.14 & 10.81 & ~1.33 \\
\hfill 3772 & ~B & \hfill $-$15.98 & \hfill  $-$9.60~~ & ~~~6 & \hfill 11.7 & ~12.22 & ~~10.79 & 12.17 & 10.94 & ~1.23 \\
\hfill 1103 & ~C & \hfill $-$41.61 & \hfill     8.22~~ & ~~~5 & \hfill 40.0 & ~12.57 & ~~11.35 & 12.52 & 11.47 & ~1.05 \\
\hfill 2851 & ~D & \hfill $-$23.58 & \hfill  $-$3.02~~ & ~~~5 & \hfill 19.4 & ~12.65 & ~~11.25 & 12.60 & 11.39 & ~1.21 \\
\hfill 6428 & ~E & \hfill    17.70 & \hfill  $-$5.21~~ & ~~~6 & \hfill 22.4 & ~12.66 & ~~11.58 & 12.61 & 11.68 & ~0.93 \\
\hfill 5610 & ~F & \hfill     0.92 & \hfill $-$22.33~~ & ~~~6 & \hfill 16.3 & ~12.67 & ~~11.54 & 12.62 & 11.65 & ~0.97 \\
\hfill  758 & ~G & \hfill $-$46.77 & \hfill    14.67~~ & ~~~5 & \hfill 47.4 & ~12.69 & ~~11.34 & 12.63 & 11.47 & ~1.16 \\
\hfill 4982 & ~H & \hfill  $-$5.95 & \hfill $-$18.11~~ & ~~~6 & \hfill 11.2 & ~12.70 & ~~11.47 & 12.65 & 11.59 & ~1.06 \\
\hfill 1203 & ~I & \hfill $-$40.26 & \hfill  $-$4.32~~ & ~~~8 & \hfill 35.8 & ~12.74 & ~~11.49 & 12.69 & 11.61 & ~1.08 \\
\hfill 4976 & ~J & \hfill  $-$5.99 & \hfill    13.09~~ & ~~~6 & \hfill 20.1 & ~12.74 & ~~11.55 & 12.69 & 11.66 & ~1.03 \\
\hfill 5407 & ~K & \hfill  $-$1.58 & \hfill     9.72~~ & ~~~6 & \hfill 17.0 & ~12.77 & ~~11.45 & 12.72 & 11.58 & ~1.14 \\
\hfill 2739 & ~L & \hfill $-$24.59 & \hfill $-$19.86~~ & ~~~7 & \hfill 23.8 & ~12.95 & ~~11.73 & 12.90 & 11.85 & ~1.05 \\
\hfill 2107 & ~M & \hfill $-$30.87 & \hfill $-$16.68~~ & ~~~7 & \hfill 28.0 & ~12.98 & ~~11.71 & 12.93 & 11.84 & ~1.09 \\
\hfill 1439 & ~N & \hfill $-$37.57 & \hfill $-$24.52~~ & ~~~7 & \hfill 37.4 & ~13.04 & ~~11.75 & 12.99 & 11.88 & ~1.11 \\
\hfill 2581 & ~O & \hfill $-$26.00 & \hfill $-$31.50~~ & ~~~7 & \hfill 32.5 & ~13.15 & ~~12.03 & 13.10 & 12.13 & ~0.97 \\
\hfill 6275 & ~P & \hfill    10.68 & \hfill $-$10.90~~ & ~~~6 & \hfill 15.8 & ~13.21 & ~~12.01 & 13.16 & 12.13 & ~1.03 \\
\hfill 2878 & ~Q & \hfill $-$23.36 & \hfill $-$10.09~~ & ~~~7 & \hfill 19.0 & ~13.22 & ~~12.10 & 13.17 & 12.20 & ~0.97 \\
\hfill 3076 & ~R & \hfill $-$21.67 & \hfill $-$46.30~~ & ~~~7 & \hfill 42.9 & ~13.22 & ~~12.09 & 13.17 & 12.20 & ~0.97 \\
\hfill 4204 & ~S & \hfill $-$12.45 & \hfill     5.10~~ & ~~~6 & \hfill 14.4 & ~13.21 & ~~12.26 & 13.17 & 12.35 & ~0.82 \\
\hfill 2204 & ~T & \hfill $-$29.76 & \hfill     6.25~~ & ~~~5 & \hfill 28.5 & ~13.31 & ~~12.20 & 13.27 & 12.31 & ~0.96 \\
\hfill 5540 & ~U & \hfill     0.00 & \hfill     0.00~~ & ~~~6 & \hfill  8.4 & ~13.41 & ~~12.08 & 13.35 & 12.20 & ~1.15 \\
\hfill 1118 & ~V & \hfill $-$41.38 & \hfill $-$17.90~~ & ~~~8 & \hfill 38.4 & ~13.47 & ~~12.39 & 13.42 & 12.49 & ~0.93 \\
\hfill 4841 & ~W & \hfill  $-$7.19 & \hfill  $-$6.16~~ & ~~~6 & \hfill  2.7 & ~13.48 & ~~12.41 & 13.43 & 12.51 & ~0.92 \\
\hfill  673 & ~X & \hfill $-$48.09 & \hfill $-$24.79~~ & ~~~8 & \hfill 47.0 & ~13.52 & ~~12.58 & 13.47 & 12.66 & ~0.81 \\
\hfill 4009 & ~Y & \hfill $-$14.22 & \hfill $-$41.45~~ & ~~~7 & \hfill 35.8 & ~13.60 & ~~12.47 & 13.56 & 12.59 & ~0.97 \\

\hline
\end{tabular}

\begin{itemize}
\item[~~~$^a$]{ID\#s from the complete version of Table~1 which is sorted by
right ascension (see text).}

\item[~~~$^b$]{Letter IDs are sorted by $V$ brightness and are shown in
Figure~1.}
\end{itemize}

\end{table}

\noindent
\begin{table}
\begin{tabular}{llllll}
\multicolumn{6}{c}{\smcap Table~2.} \\
\multicolumn{6}{c}{A Comparison of Blue Straggler Frequencies} \\
%\multicolumn{6}{c}{} \\
\hline
\hline
\multicolumn{1}{l}  {Cluster} &
\multicolumn{1}{c} {$F_{\rm BSS}$} &
\multicolumn{2}{c} {Region~Studied$^a$} &
\multicolumn{1}{c} {$\rho_{0}^a$} &   
\multicolumn{1}{l}  {Source} \\

 & & & \multicolumn{1}{c} {[pc]} & \multicolumn{1}{c} {[$M_\odot\,\rm
 pc^{-3}$]} & \\

\hline

M13 & $0.04$--$0.07$ & $r\lesssim1.7\,r_{\rm core}$ & \hfill 2.3 & \hfil
$2.5\times10^3$ \hfill & This paper \\

\hline

M15 & $0.06\pm0.01$ & $r\lesssim15\,r_{\rm core}^b$ & \hfill 1.7 & \hfil
$1.6\times10^6$ \hfill & Guhathakurta et~al.\ (1996) \\

M30 & $0.19\pm0.04$ & $r\lesssim10\,r_{\rm core}^b$ & \hfill 0.7 & \hfil
$8\times10^5$ \hfill & Yanny et~al.\ (1994b) \\

47~Tuc & $0.07\pm0.01^c$ & $r\lesssim3\,r_{\rm core}$ & \hfill 1.6 &
\hfil $1.3\times10^5$ \hfill & Paper~I \\

M3 & $0.09\pm0.02^d$ & $r\lesssim0.7\,r_{\rm core}$ & \hfill 1.0 & \hfil
$3.2\times10^3$ \hfill & Paper~III \\

NGC~5053 & $0.14\pm0.03$ & $r<1\,r_{\rm core}$ & \hfill 10 & \hfil 4
\hfill & Nemec \& Cohen (1989) \\

M3$^e$ & $0.09\pm0.03$ & $r\sim15\,r_{\rm core}$ & \hfill 20 &
\hfil ~~1$^e$ \hfill & Paez et~al.\ (1990) \\

\hline
\end{tabular}

\begin{itemize}
\item[~~~$^a$]{Central density and distance: Pryor \& Meylan (1993); core
radius: Djorgovski (1993).}

\item[~~~$^b$]{The core radius is poorly defined for the post core collapse
clusters M15 and M30; we arbitrarily assume $r_{\rm core}=2''$, the rough
upper limit to their core radii.}

\item[~~~$^c$]{Corrected for the 50\% BSS detection efficiency in our $V$-
and $I$-band WF/PC-I data.}

\item[~~~$^d$]{Bolte et~al.'s (1993) ground-based study of the core of M3
found a BSS frequency of only 0.04, but their data were shown to be
incomplete (Paper~III).}

\item[~~~$^e$]{The space density of stars in this outer M3 field is expected
to be about $3\times10^{-4}$ of the central value.}
\end{itemize}

\end{table}

\bigskip\bigskip
\bigskip\bigskip

\begin{table}
\begin{tabular}{lllll}
\multicolumn{5}{c}{\smcap Table~3.} \\
\multicolumn{5}{c}{Radial Distribution of Evolved Stellar Populations in M13} \\
%\multicolumn{5}{c}{} \\
\hline
\hline

\multicolumn{1}{l} {} & \multicolumn{4}{c} {Limiting~Radii} \\

\cline{2-5}

\multicolumn{1}{l} {Stellar Subsample} &
\multicolumn{1}{c} {{\it r}$<$15$\farcs$0} &
\multicolumn{1}{c} {15$\farcs$0$\leq${\it r}$<$25$\farcs$8} &
\multicolumn{1}{c} {25$\farcs$8$\leq${\it r}$<$37$\farcs$6} &   
\multicolumn{1}{c} {37$\farcs$6$\leq${\it r}$<$65$\farcs$6} \\

\hline

Bright stars ($V<18.0$)$^a$
& \hfil 233 \hfill & \hfil 235 \hfill & \hfil 234 \hfill & \hfil 234 \hfill \\
All~stars ($V\lesssim20$)
& \hfil 643 \hfill & \hfil 668 \hfill & \hfil 731 \hfill & \hfil 835 \hfill \\
Faint RGB/subgiants
& \hfil 180 \hfill & \hfil 180 \hfill & \hfil 169 \hfill & \hfil 176 \hfill \\
Bright RGB/AGB
& \hfil 32 \hfill & \hfil 29 \hfill & \hfil 34 \hfill & \hfil 27 \hfill \\ 
HB
& \hfil 13 \hfill & \hfil 22 \hfill & \hfil 24 \hfill & \hfil 25 \hfill \\ 
BSS (all candidates)
& \hfil 8 \hfill & \hfil 4 \hfill & \hfil 7 \hfill & \hfil 6 \hfill \\

\hline

\end{tabular}
\medskip

\begin{itemize}
\item[~~~$^a$]{The sample should be complete to this limiting magnitude; the
limiting radii were chosen to include the same number of bright stars in each
radial bin.}
\end{itemize}

\end{table}

\vfill \eject

\vglue 6in

\begin{figure}
%Fig. 1
%\plotone{Cohen.fig1.ps}
% \vskip 7.0truein
\caption{Figure 1, which is listed separately and is in jpeg format,
is a negative (color-reversed) {\it HST\/} F555W image of the core of
M13.  The PC mosaic consists of CCDs PC5--PC8, with PC5 in the upper
right and continuing counterclockwise.  The gaps between the four CCD
frames are represented approximately to scale.  A selected set of
bright, relatively isolated reference stars are identified by the
letters A--Y; the positions and brightnesses of these reference stars
are listed in Table~1.  The ``+'' near the center of PC6 marks the
cluster centroid and the density of stars is highest on PC6.  The
scale and orientation of the image are indicated (vertical is
$38\fdeg46$ East of North).}
\label{fig1.fig}
\end{figure}

\begin{figure}
\plotone{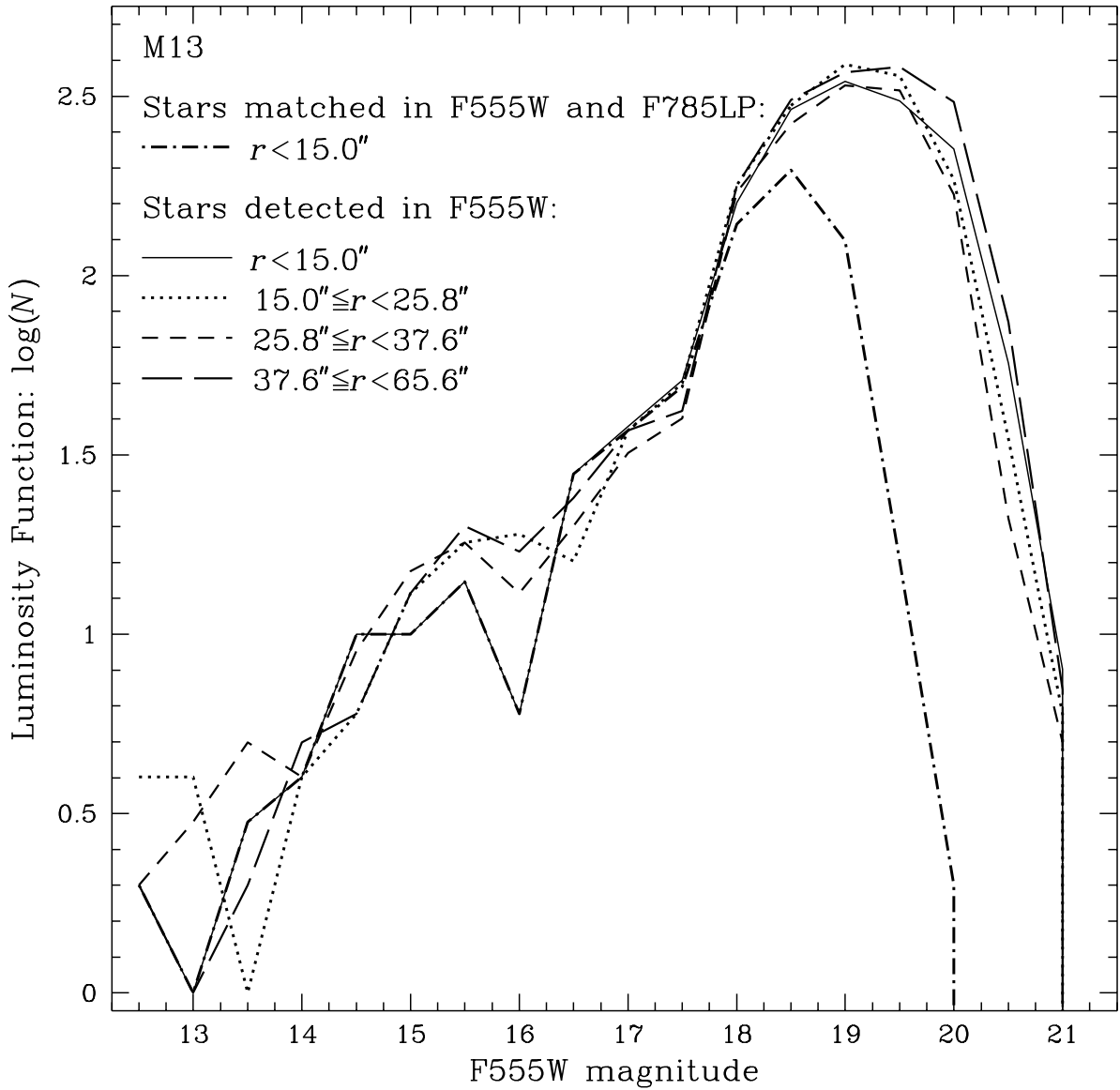}
%Fig. 2
\caption{The observed stellar luminosity function of M13 in the F555W band
(roughly similar to Johnson $V$) in four concentric radial bins.  The
rollover beyond $\rm F555W\sim19$ is caused by incompleteness.  The bold
dot-dashed line represents stars in the innermost radial bin that are
detected in both F555W and F785LP bands; incompleteness sets in about 1~mag
brighter in this sample than in the corresponding F555W-only sample.  The
weak bump at $\rm F555W\sim15$--16 is caused by stars in the extended blue
HB.}
\label{fig2.fig}
\end{figure}

\begin{figure}
\plotone{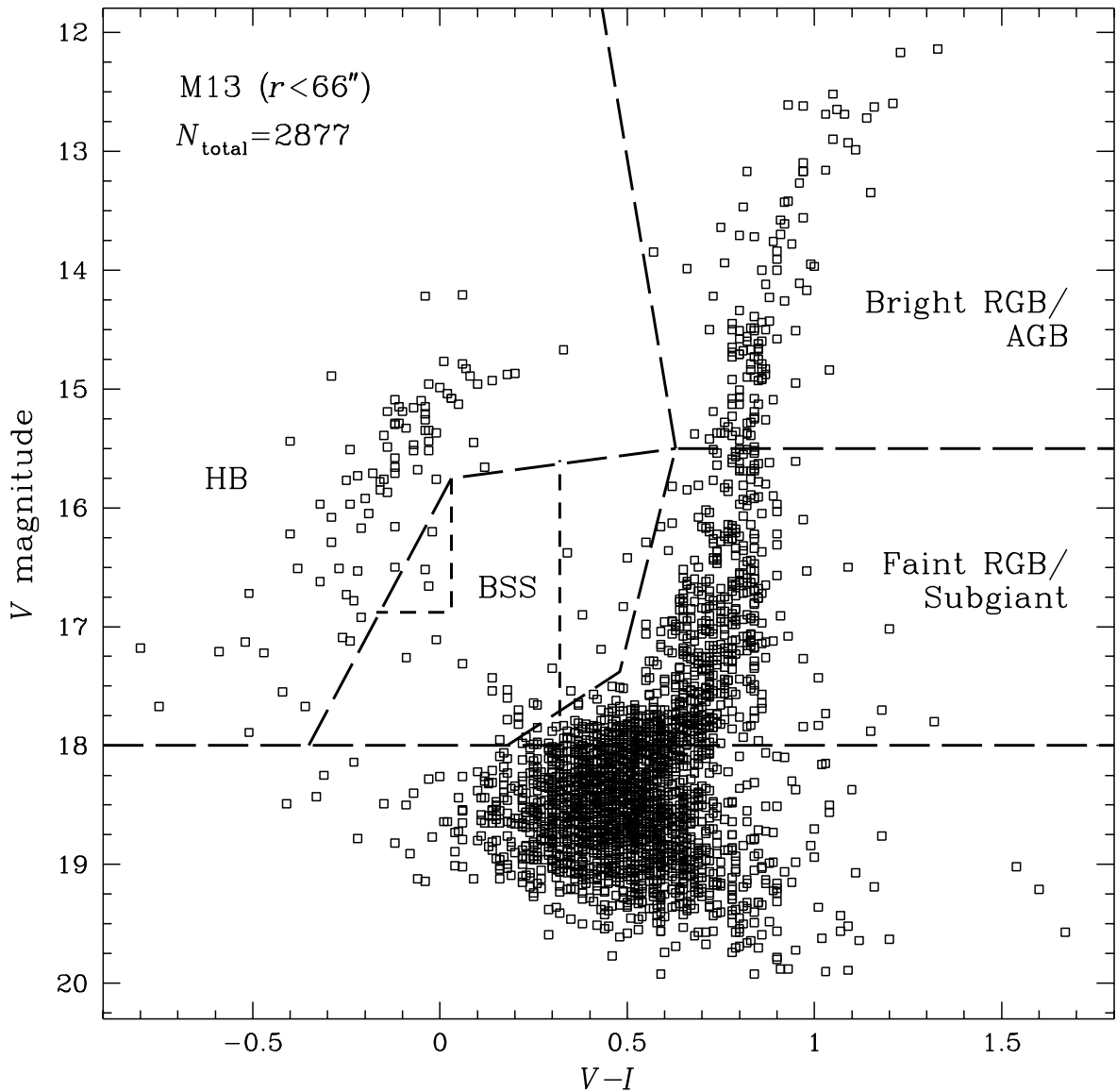}
%Fig. 3
\caption{A color--magnitude diagram of all 2877~matched F555W and F785LP
detections in M13 ($r<66^{\prime\prime}$).  The dashed lines indicate the
boundaries that have been used to distinguish between various stellar types.
The dotted lines in the blue straggler star (BSS) region enclose the
10~``possible'' BSS candidates.  The sample is increasingly incomplete
fainter than $V\sim18.5$, below the main sequence turnoff.}
\label{fig3.fig}
\end{figure}

\begin{figure}
\plotone{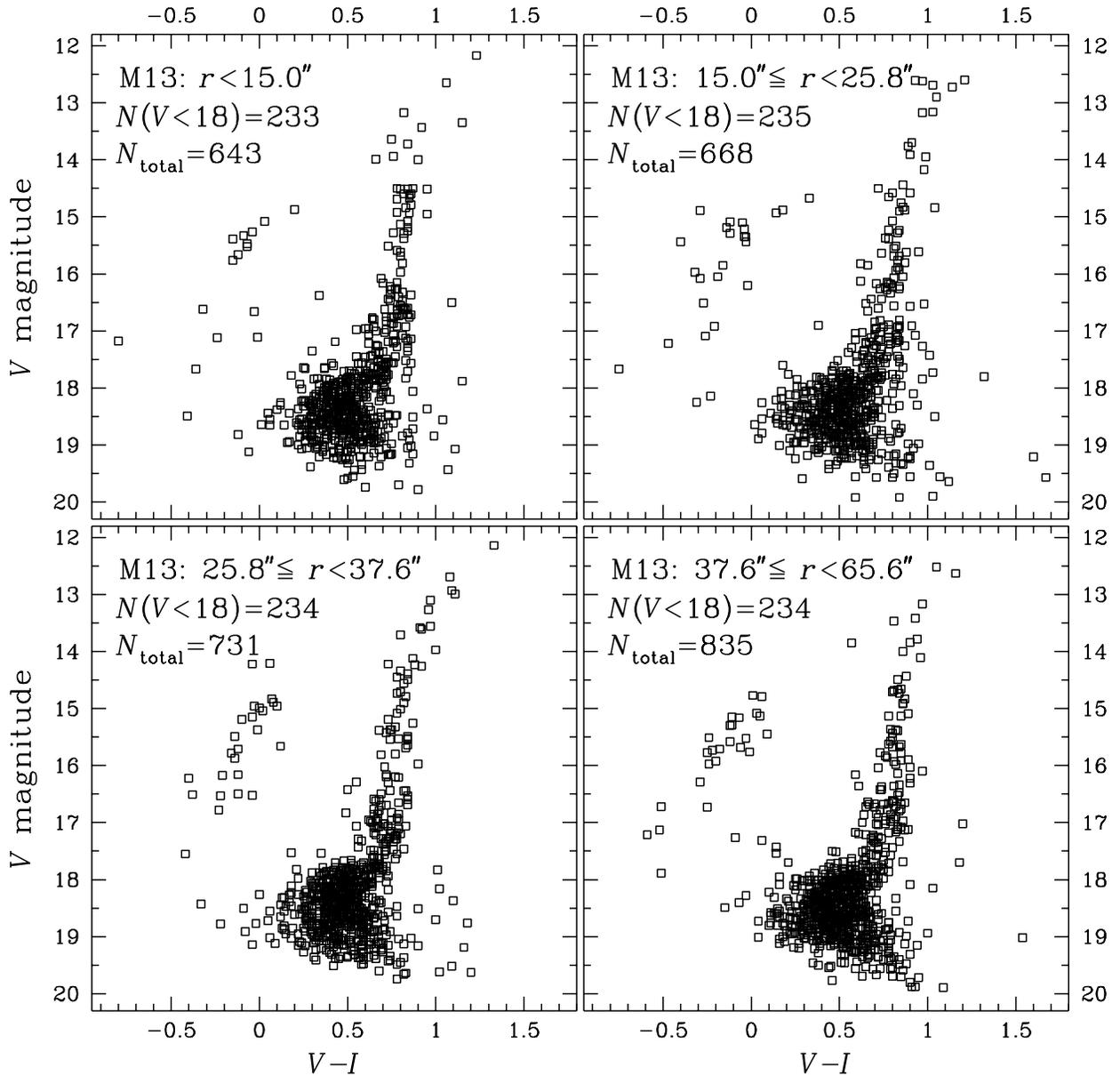}
%Fig. 4
\caption{A composite of $V$ vs $V-I$ color--magnitude diagrams in four radial
bins around the approximate center of M13.  The limiting radii have been
chosen to ensure that the bins contain equal numbers of stars with $V<18$
(the sample is complete down to this limiting magnitude).  The degree of
completeness for faint ($V\gtrsim18$), turnoff stars is somewhat higher in
the outer bins (3 \& 4) than in the inner bins.  The radial bins do {\it
not\/} represent entire annuli.  The innermost bin appears to be slightly
deficient in horizontal branch stars relative to the bins further out.}
\label{fig4.fig} 
\end{figure}

\begin{figure}
\plotone{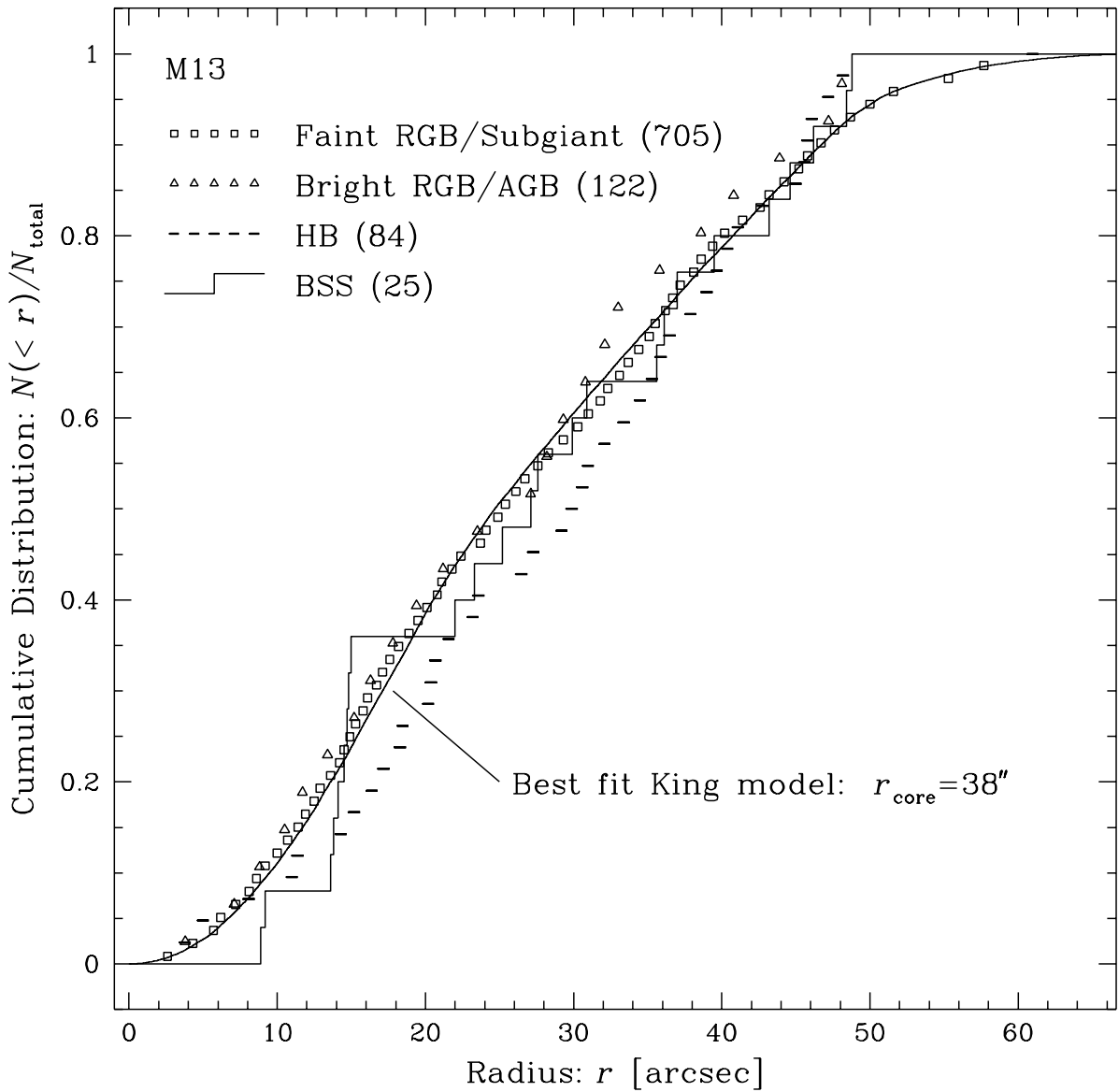}
%Fig. 5
\caption{The cumulative radial distribution of different kinds of post main
sequence stars in M13.  Only every second HB star, every fifth bright RGB/AGB
star, and every tenth faint RGB/subgiant star is shown for the sake of
clarity.  The smooth solid line shows an $r_{\rm core}=38^{\prime\prime}$
King profile; this is consistent with the radial distribution red giants to
within Poisson errors.  The numbers within parentheses indicate the number of
stars of each type.  There is a suggestion that the HB stars have a weaker
degree of central concentration than the giants.}
\label{fig5.fig}
\end{figure}

\begin{figure}
\plotone{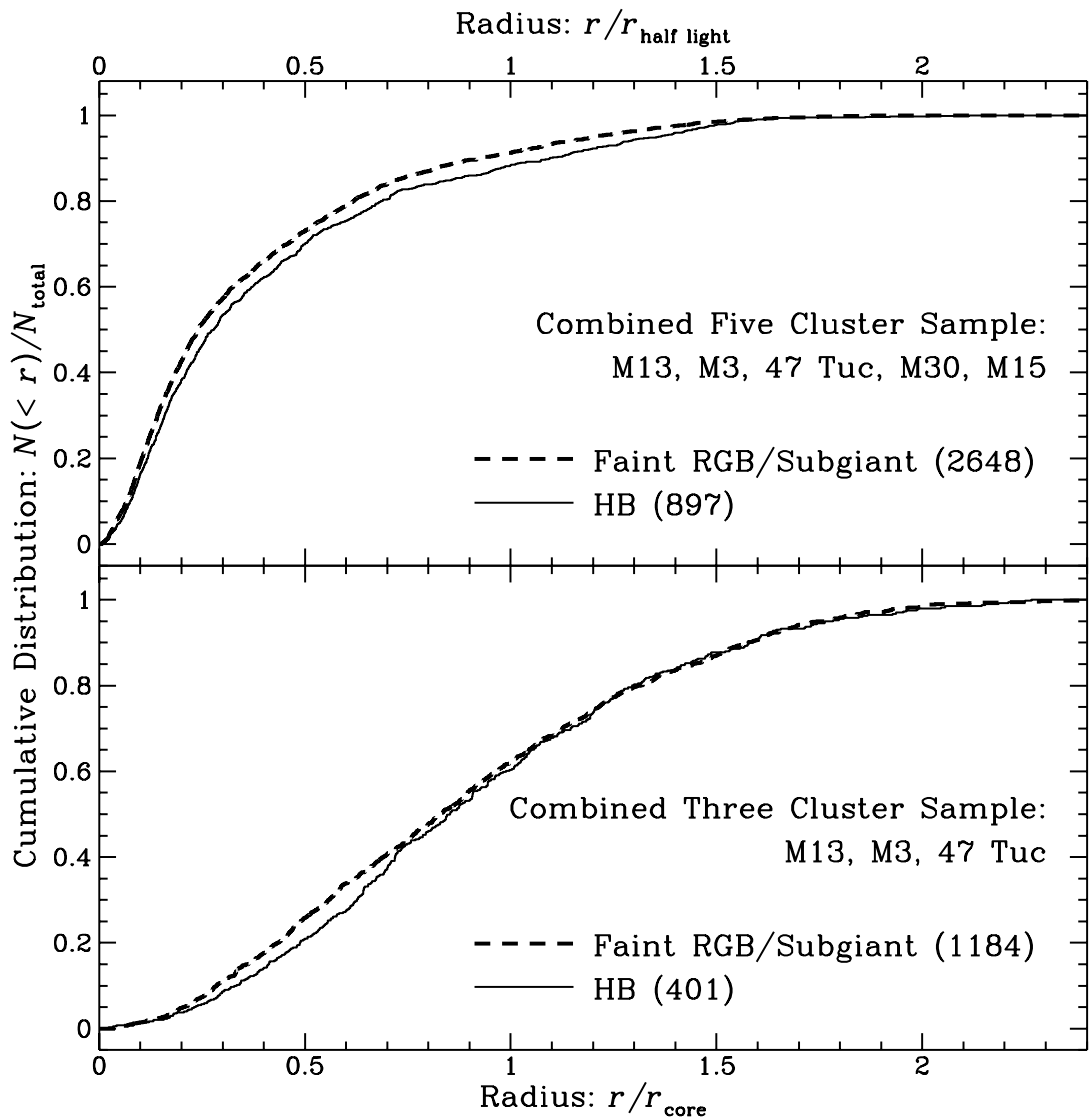}
%Fig. 6
\caption{The cumulative radial distribution of HB stars (solid line)
compared to that of faint RGB/subgiant stars (dashed line) in combined
samples of globular clusters.  The number of stars in each category is shown 
within parentheses.~~~
{\bf Top:}~Combined data from M13, M3, 47~Tuc, M30, and M15, with the
projected radius normalized by the half light radius.  The HB stars are less
centrally concentrated than the giants at the 97\% ($\sim2\sigma$)
significance level.~~~
{\bf Bottom:}~Combined data from only the three King model clusters, 
M13, M3, and 47~Tuc, with the projected radius normalized by the core radius.
The HB stars are less centrally concentrated than the giants at only the
85\% ($\sim1.5\sigma$) significance level.}
\label{fig6.fig}
\end{figure}

\begin{figure}
\plotone{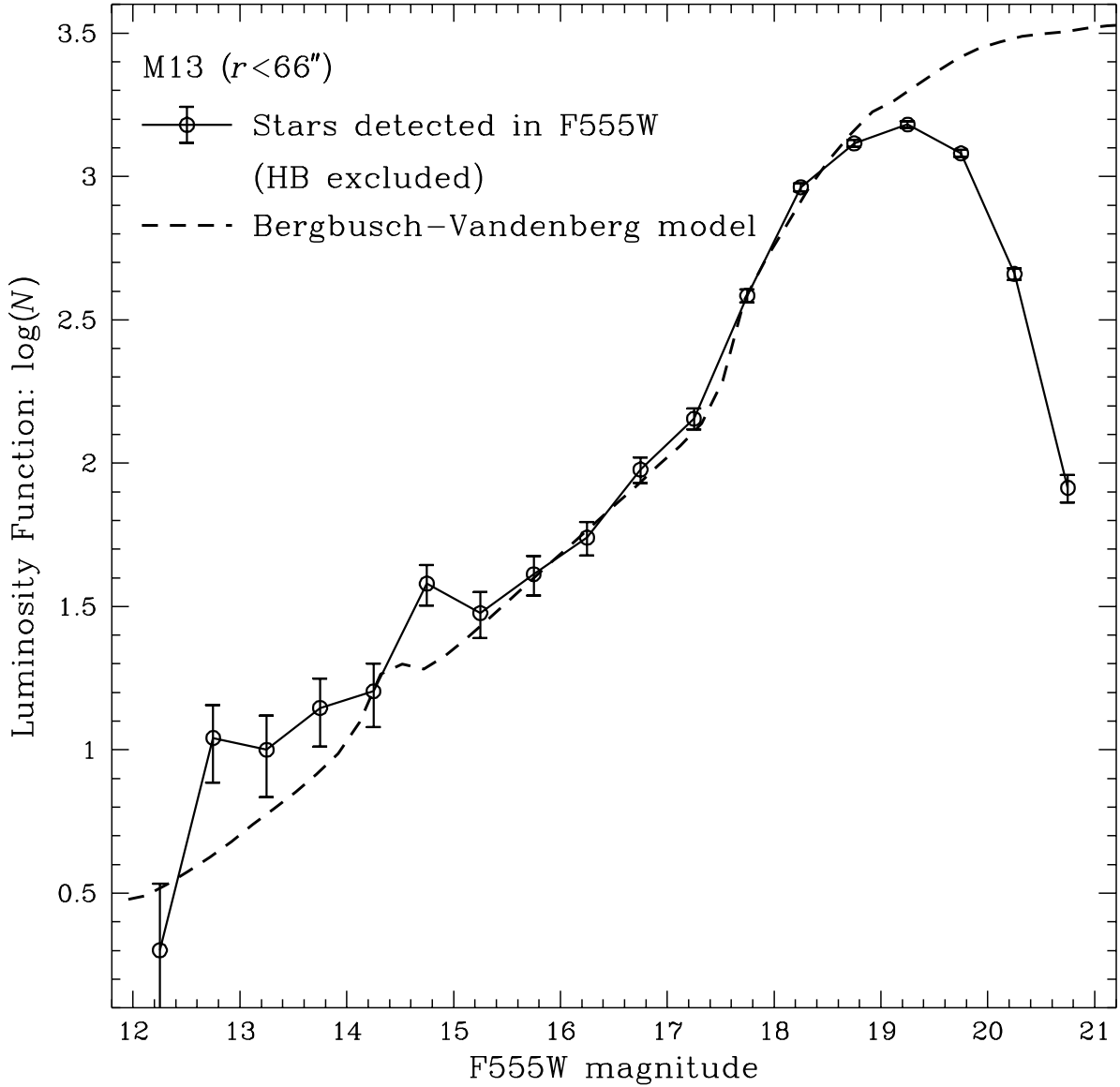}
%Fig. 7
\caption{The observed stellar luminosity function of M13 (points) compared to
the Bergbusch \& Vandenberg (1992) model for an old, metal-poor population
(dashed line), corrected for the distance modulus and line-of-sight
extinction of M13.  The error bars indicate the $1\sigma$ Poisson error in
0.5~mag bins.  The model is normalized to the observed number of stars with
$\rm F555W=19$, adjusted for 90\% completeness.  It is a good match to the
shape of M13's LF over the range $\rm F555W\approx15$--18.5, beyond which the
M13 sample is incomplete.  Note the significant excess of bright RGB stars
($\rm F555W\approx12.5$--15) in M13 relative to the model.}
\label{fig7.fig}
\end{figure}

\end{document}